\documentclass[conference]{IEEEtran}
\IEEEoverridecommandlockouts
\usepackage{cite}
\usepackage{amsmath,amssymb,amsfonts}
\usepackage{algorithmic}
\usepackage{graphicx}
\usepackage{textcomp}
\usepackage{xcolor}
\usepackage{subfiles}
\usepackage{tcolorbox}
\usepackage{url}
\def\BibTeX{{\rm B\kern-.05em{\sc i\kern-.025em b}\kern-.08em
    T\kern-.1667em\lower.7ex\hbox{E}\kern-.125emX}}
\begin{document}

\title{Evaluating Large Language Models for Code Review}

\author{\IEEEauthorblockN{ Umut Cihan}
\IEEEauthorblockA{\textit{Bilkent University} \\
Ankara, Turkey \\
umut.cihan@bilkent.edu.tr }
\and
\IEEEauthorblockN{Arda İçöz}
\IEEEauthorblockA{\textit{Bilkent University} \\
Ankara, Turkey \\
arda.icoz@bilkent.edu.tr}
\and
\IEEEauthorblockN{Vahid Haratian}
\IEEEauthorblockA{\textit{Bilkent University} \\
Ankara, Turkey \\
vahid.haratian@bilkent.edu.tr}
\and
\IEEEauthorblockN{Eray Tüzün}
\IEEEauthorblockA{\textit{Bilkent University}\\
Ankara, Turkey \\
eraytuzun@cs.bilkent.edu.tr}}

\maketitle

\begin{abstract}
Context: Code reviews are crucial for software quality. Recent AI advances have allowed large language models (LLMs) to review and fix code; now, there are tools that perform these reviews. However, their reliability and accuracy have not yet been systematically evaluated.

Objective: This study compares different LLMs’ performance detecting code correctness and suggesting improvements.

Method: We tested GPT-4o and Gemini 2.0 Flash on 492 AI-generated code blocks of varying correctness, along with 164 canonical code blocks from the HumanEval benchmark. To simulate the code review task objectively, we expected LLMs to assess code correctness and improve the code if needed. We ran experiments with different configurations and reported on the results.

Results: With problem descriptions, GPT4o and Gemini 2.0 Flash correctly classified code correctness 68.50\% and 63.89\% of the time, respectively, and corrected the code 67.83\% and 54.26\% of the time for the 492 code blocks of varying correctness. Without problem descriptions, performance declined. The results for the 164 canonical code blocks differed, suggesting that performance depends on the type of code. 


Conclusion:  LLM code reviews can help suggest improvements and assess correctness, but there is a risk of faulty outputs. We propose a process that involves humans, called the "Human-in-the-loop LLM Code Review,"  to promote knowledge sharing while mitigating the risk of faulty outputs.
\end{abstract}


\section{Introduction}

Code review is a crucial software development practice that enhances code quality, facilitates knowledge sharing, and detects defects \cite{Macleod2018_CRInTrenches, Baum2016_ClassificationSchemeForIndustrialCR}. Formal inspections, a longstanding form of code review \cite{Fagan1976_codeInspections, Ciolkowski_2022}, require practitioners to examine and modify code changes before they are merged into production \cite{Davila2021_literatureReview}.

Code review, reliant on manual effort and a practitioner's expertise \cite{Shull2008_historyOfInspections}, is labor-intensive and often challenges practitioners with time allocation \cite{Kononenko2016_CRQualityHowDevelopersSee, Macleod2018_CRInTrenches}. Modern industry practices address this by adopting a lightweight, tool-based, informal process known as modern code review \cite{Bacchelli2013_modernCodeReview, Davila2021_literatureReview}. Considering the current challenges, automation can offer significant time savings and prevent shallow reviews. Recent AI advancements have led to several AI-assisted code review tools \cite{qodo}, \cite{CodeRabbit}, \cite{GithubCopilot} powered by large language models (LLMs) that reduce manual effort and streamline project schedules. However, empirical analysis of the quality of their reviews is crucial for determining their reliability and accuracy.

Our study aims to illuminate LLM capabilities in code reviews. Reviewers first assess code correctness in change requests and then suggest improvements if issues arise. We developed a methodology and two research questions addressing these aspects.
We aim to answer the following research questions. 
\begin{description}
    \item [RQ1:] How accurately can large language models (LLMs) evaluate code changes for approval or rejection?
    \item [RQ2:] How effective are the code improvement suggestions generated by large language models (LLMs) in improving code correctness?
\end{description}

We developed a setup that uses code blocks along with their unit tests. The setup prompts the LLM to assess code correctness and suggest improvements for each code block. We evaluate the LLM’s assessment against unit test results and test its suggestions with the same unit tests. A suggestion is a correction if the new code passes all unit tests. In code reviews, authors often include comments. To reflect this, we added the problem description to some prompts and omitted it from others, then tested both and reported the results.

Our experiments led to several key findings. Firstly, the results indicated that LLMs would be unreliable in a fully automated code review environment. Secondly, incorporating problem descriptions into prompts consistently improved performance, highlighting the importance of code comments and pull request descriptions. Finally, our results varied across different datasets. This underlined the need for custom testing tailored to the target codebase. We shared our experiment setup and source code to support practitioners \footnote{https://doi.org/10.5281/zenodo.14962566}.

Based on our findings, we propose a process incorporating human oversight instead of relying solely on complete automation, that is  "Human-in-the-loop LLM Code Review." The process involves LLMs reviewing all change requests, while a human "Review Responsible" decides whether there is a need for human review. This process would resolve reliability issues and decrease the need for manual effort. It also allows for knowledge sharing, an essential aspiration for code review \cite{Macleod2018_CRInTrenches}. The process can be tailored according to organizational needs. Our suggested process and experiment setup enable practitioners to implement their LLM-assisted code review processes effectively.

In Section \ref{background}, the reader can find background information on the code review process and LLMs alongside related work. In Section \ref{methodology}, we describe our methodology. In Section \ref{results}, readers can find results from our experiments. The results are discussed in section \ref{discussion}, and future research directions are given. In Section \ref{threatstovalidity}, we list threats to validity. In Section \ref{conclusion}, we conclude our study.




\section{Background}
\label{background}
\subsection{Code Review}
\label{Code_Review}
Developers widely use code reviews to inspect changes before integration \cite{Davila2021_literatureReview}. Formal inspections, introduced by Fagan in 1976 \cite{Fagan1976_codeInspections}, boosted productivity and quality but were often too time-consuming for universal adoption, depending on organizational context \cite{Shull2008_historyOfInspections}. By 2013, Modern Code Review (MCR) emerged as a lightweight, informal, tool-based alternative \cite{Bacchelli2013_modernCodeReview}.

MCR is now prevalent in companies like Google \cite{Sadowski2018_McrAtGoogle}, AMD \cite{Rigby2013_convergentContemporaryReview}, and Microsoft \cite{Rigby2013_convergentContemporaryReview}, as well as in OSS (open-source software) \cite{Beller2014_McrInOSS}. A 2018 Google study \cite{Sadowski2018_McrAtGoogle} highlighted its importance for codebase understanding, integrity, readability, and consistency. While review coverage significantly impacts quality, it's not the sole factor, and poor reviews can harm software quality \cite{McIntosh2015_impactOfModernCodeReview}.

However, code review is time-consuming. A 2013 survey found OSS participants spent an average of 6.4 hours per week on reviews \cite{Bosu2013_ImpactOfPeerCR}. Similarly, a 2018 Microsoft study identified timely feedback and time management as major challenges for developers \cite{Macleod2018_CRInTrenches}.

\subsection{Large Language Models (LLMs)}
\label{LLM}
Natural language processing has advanced significantly in recent years. In 2014, the introduction of Long Short-Term Memory (LSTM) networks by Sutskever et al. \cite{Sutskever2014_SeqToSeqNeuralNetworks} and recurrent neural networks (RNNs) by Cho et al. \cite{Cho2014_RNN} led to models that handle sequential data (e.g., text) more effectively than previous approaches. The Transformer architecture, introduced by Vaswani et al. \cite{Vaswani2017_Transformers} in 2017, further advanced the field with self-attention mechanisms that assess word importance without relying on their distance. Google introduced BERT \cite{Devlin2018_BERT} in 2018, shifting the field toward pre-trained models. BERT and successors like GPT \cite{Radford2018_GPT1}, RoBERTa \cite{Liu2019_RoBERTa}, and T5 \cite{Raffel2019_T5} demonstrated that models pre-trained on large text corpora can be fine-tuned for specific tasks with minimal additional data.

OpenAI's GPT-2 \cite{Radford2019_GPT2} (2019) and GPT-3 \cite{Brown2020_GPT3} (2020) showcased LLM versatility. In 2021, Codex \cite{Chen2021_Codex}, fine-tuned for programming, led to GitHub Copilot \cite{GithubCopilot}. ChatGPT \cite{ChatGPT} and GPT-4 \cite{Achiam2023_GPT4} further boosted LLM use in programming. In March 2024, Anthropic AI introduced the Claude 3 family \cite{Claude3_2024}, with Claude 3 Opus outperforming state-of-the-art LLMs. In May 2024, OpenAI unveiled GPT-4o \cite{GPT-4o_2024}, a faster, cross-modal variant of GPT-4 that excels across benchmarks. In December 2024, Google introduced the Gemini 2.0 family, outperforming their models of the previous generation \cite{Gemini}. 

\subsection{Related Work}
\label{related_work}

Automating code reviews is motivated by evidence of the process's time-consuming nature \cite{Bosu2013_ImpactOfPeerCR, Macleod2018_CRInTrenches}. Most efforts have focused on recommending suitable reviewers \cite{AlZubaidi2020_WorkloadAwareReviewerRecommendation, Asthana2019_WhoDoReviewerSuggestion, Jiang2019_WhoShouldMakeDecisionOnThisPR, Ying2016_EARec, Mirsaeedi2020_MitigatingTurnoverCRRecommendation, Ouni2016_SearchBasedRevRec, Rahman2016_Correct, Thongtanunam2015_WhoShouldReview}.

To improve efficiency, the review process itself can be automated as a code-to-comment task \cite{Tufano2021_TowardsAutomatingCR}. In 2018, Gupta and Sundaresan \cite{Gupta2018_IntelligentCRUsingDL} introduced a deep learning model that matched code blocks with historical reviews. In 2019, Li et al. \cite{Li2019_DeepReview} proposed a CNN-based model to predict change approval, while Shi et al. \cite{Shi2019_AutoCRByLearningRev} used a CNN-LSTM framework for the same purpose. In 2022, Hong et al. \cite{Hong2022_CommentFinder} presented CommentFinder, which leverages information retrieval to suggest code comments, and Li et al. \cite{Li2022_AUGER} introduced AUGER, a system that automatically generates review comments using a pre-trained model. Tufano et al. \cite{Tufano2022_UsingPreTrainedModelsForCR} employed a T5 model \cite{Raffel2019_T5} for code review automation. Thongtanunam et al. \cite{Thongtanunam2022_AutoTransform} developed a model to modify source code automatically during reviews to reduce manual effort, while a tool with the same purpose was found useful at Google \cite{froemmgen_2024}. Li et al. \cite{Li2022_AutomatingCRByLargeScalePreTraining} further explored automation via large-scale pre-training on diverse code datasets. Zhou et al. \cite{Zhou2023_GenerationBasedCRAutomation} introduced the Edit Progress (EP) metric to capture partial progress in automated reviews. Rasheed et al. \cite{Rasheed2024_AIPoweredCR} and Tang et al. \cite{Tang_2024} developed LLM agents to automate code review.

In 2024, Tufano et al. \cite{Tufano2024_CodeReviewAutomation} qualitatively evaluated prior work \cite{Tufano2022_UsingPreTrainedModelsForCR, Hong2022_CommentFinder, Li2022_AutomatingCRByLargeScalePreTraining} alongside ChatGPT \cite{ChatGPT}. Although the ChatGPT version was unspecified, findings indicated it could serve as a competitive baseline for the comment-to-code task (i.e., revising code after a review), though it did not outperform state-of-the-art methods in the code-to-comment task \cite{Tufano2024_CodeReviewAutomation}. Unlike prior work, our study examines LLMs as code approvers, responsible for making change merge decisions and offering suggestions. We established an experimental setup for benchmarking various models, enabling practitioners to identify the optimal model and prompt configuration for their specific codebase.

\section{Methodology}
\label{methodology}

Developers have different expectations about what makes a “good” code review \cite{Bacchelli2013_modernCodeReview}. To ensure a robust evaluation, we focus on the merge approval decision, which governs the procedure for integrating change requests into the mainline code.   This decision is based on code correctness, defined as the ability of the code to perform its intended functionality in all cases. When new code is submitted, reviewers determine whether it should be merged into the mainline code; if rejected, they offer suggestions for improvement. Similarly, we expect LLMs to deliver a verdict on code correctness and offer suggestions when necessary. 

Our dataset comprises code blocks with unit tests, offering an objective standard for code correctness and a means to assess the effectiveness of improvement suggestions. Code blocks that pass all unit tests are deemed "Correct," while those that do not are considered "Incorrect."

\subsection{Dataset}
We use the HumanEval dataset \cite{chen2021_HumanEvalDataset}, a popular dataset for evaluating LLMs with interview-level code generation questions, alongside LLM-generated code blocks from Yetistiren et al. \cite{Yetistiren2023_evaluateCodeQualityOfAI} for solving HumanEval questions. Our two datasets are 164 canonical solutions from the HumanEval dataset that pass all unit tests and 492 AI-generated code blocks from three tools (ChatGPT 9 Jan '23, Amazon CodeWhisperer Jan '23, and GitHub Copilot v1.70.8099), all written in Python. The datasets are referred to as "Ground Truth" and "Mixed" respectively. The 492 AI-generated blocks are categorized as 234 "Correct" and  258 "Incorrect."  We chose this dataset for its unit tests and its diversity of correctness.

\subsection{Test Setup}
\begin{figure}[h]
    \centering
    \includegraphics[width=1\linewidth]{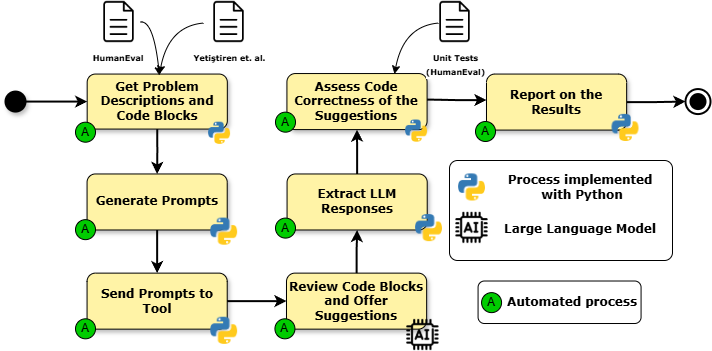}
    \caption{Test Setup}
    \label{fig:methodology_test_setup}
\end{figure}

In our test setup, we simulate a code review scenario where reviewers must approve or reject a proposed change. When rejecting a change, reviewers typically suggest improvements. Therefore, our prompt instructs the LLM to classify the code as correct or "Incorrect" and to provide a code suggestion. The detailed steps to our methodology can be seen in Figure \ref{fig:methodology_test_setup}.

Prompts can enhance and refine the LLM's capabilities \cite{White2023_PromptEng}. We optimized our prompt using a chain-of-thought style, a widely used prompting method. Our prompt template is given in Figure \ref{fig:prompt-template}. 

Code blocks often include descriptions that explain the code in the form of code comments or pull request descriptions. To mirror this, we use the HumanEval problem descriptions in our prompts. Since such descriptions are not always provided, we had two prompt types. In Figure \ref{fig:prompt-template}, the text in red appears only in the prompts that include problem descriptions while the text in black is the same for each prompt type.

\begin{figure}[ht]
\centering
\begin{tcolorbox}[colback=black!5!white, colframe=black!75!black]
You are an experienced Python developer. You will be provided with a code block \textcolor{red}{(and a corresponding problem description)}. Your task is to understand the intended functionality, review the code, and generate feedback. Please follow these steps:
Analyze the Code: Understand what the code is meant to do.
Consider Edge Cases: Reflect on all scenarios in which the code should operate.
Evaluate Functionality: Determine whether the code successfully implements the intended functionality across these cases.
Classify the Code:
If the code requires changes, set \texttt{classified\_type} to Incorrect.
If the code does not require any changes, set \texttt{classified\_type} to Correct.
Suggest a Corrected Code Block: In the \texttt{complete\_code} field, suggest a corrected code block if the \texttt{classified\_type} is Incorrect; otherwise, you should return the code without changing it. 

Important: Use only the classifications Correct or Incorrect (case sensitive).

Now, please review the code based on the following code block:

\textbf{\#code block}

\textcolor{red}{(And the following problem description:
\textbf{\#problem description})}

You need to respond in the following format. This is a strict requirement. 

Example output:

feedback:
  classified\_type: 
    No
    
code:
  complete\_code: 
    No 
    
Apply the following rules strictly:

- Answer should be a valid YAML, and nothing else. Do not add your thought process or any other text.

- Replace "No" values with your suggestions.

- Be careful about the indentation and syntax of your suggested code block, make sure it can run without problems.
\end{tcolorbox}
\caption{Prompt Template}
\label{fig:prompt-template}
\end{figure}
The LLM's output provides two distinct pieces of information for our evaluation. The first is the code correctness classification, indicating whether the code is "Correct" or "Incorrect". The second is the code improvement suggestion: a revised code block that should perform better (or remain unchanged if the code is "Correct") than the original.

\subsection{Evaluation Metrics}

We expect reviewers to detect code correctness accurately. To quantify this, we define "Correctness Accuracy" as the proportion of correctness assessments that match unit test results, as seen in equation \ref{eq:correctness_accuracy}. When we classify "Correct" as positive and "Incorrect" as negative, correctness accuracy is equivalent to the model accuracy. Using this classification, we also calculate false positive and false negative rates.

\begin{itemize}
    \item 
    \textbf{Correctness (Model) Accuracy:}
    \begin{equation}
\\
\label{eq:correctness_accuracy}
\frac{\text{The Number of Accurately Assessed Code Blocks}}{\text{The Number of All Code Blocks}}
\end{equation}
\end{itemize}

For effective code improvements, suggestions must pass all unit tests. We define the "Correction Ratio" as the proportion of suggestions that meet this criterion, as seen in equation \ref{eq:correction_Ratio}. 

\begin{itemize}
\item \textbf{Correction Ratio:} \\ \begin{equation}
\label{eq:correction_Ratio}
 \frac{\text{The Number of Correct Code Suggestions}}{\text{The Number of Incorrect Code Blocks}}
\end{equation}
\end{itemize}

We should also consider negative scenarios to assess the impact of code suggestions. A suggestion might worsen a code block—turning "Correct" code into "Incorrect" code. We refer to such instances as "Regressions" and define a "Regression Ratio.", as seen in equation \ref{eq:regression_Ratio}. 

\begin{itemize}
    \item 
    \textbf{Regression Ratio}\\
    \begin{equation} \label{eq:regression_Ratio}
  \frac{\text{The Number of Incorrect Code Suggestions}}{\text{The Number of Correct Code Blocks}}
\end{equation}
\end{itemize}

\subsection{Variables}
Our test setup comprises three variables. The first is the LLM: we evaluated OpenAI's GPT-4o (model version: 2024-11-20), used by code review tools like Qodo \cite{qodo} and CodeRabbit \cite{CodeRabbit}, making it a relevant candidate. We also used Google's Gemini 2.0 Flash, which, during our experiments, was the most capable API-accessible model from the Gemini 2.0 family (a competitor to GPT-4o).

The second variable is the presence of a problem description (e.g., comments or pull request descriptions). This allows us to assess how such contextual information affects LLM performance.

The third variable is the dataset. We used a mixed dataset of 492 code blocks and the HumanEval dataset's canonical solutions (ground truth dataset). These canonical solutions act as a control group, providing insights into LLM reliability. For this ground truth data, regression ratios are important, as suggestions should not worsen correct code.
With eight distinct test configurations arising from three variables, we can infer the generalizability of our findings. 


\section{Results}
\label{results}
We evaluated two state-of-the-art LLMs, Gemini and GPT-4o, using the Gemini-2.0-Flash and gpt-4o-2024-11-20 versions with the default model parameters. To ensure reliability, we ran each experiment configuration three times and reported the average results. The standard deviations ranged from 0.35\%  to 1.61\%  for correctness accuracy, from  1.02\% to 1.93\% for false positive rates, from 0.65\% to 1.07\% for false negative rates, from 0\% to 2.88\%  for regression ratios, and from 0.38\% to 1.34\% for correction ratios. A chi-square test for variance (using a 5\% threshold, df=2, and a critical value of 5.991 at $\alpha=0.05$) showed that all chi-square statistics were below the threshold. This confirms the consistency of our results, as the observed standard deviations are not statistically significant.
We share our data and code in our replication package.

\subsection{Mixed Dataset Experiment Results}
\begin{figure}[h]
    \centering
    \includegraphics[width=1\linewidth]{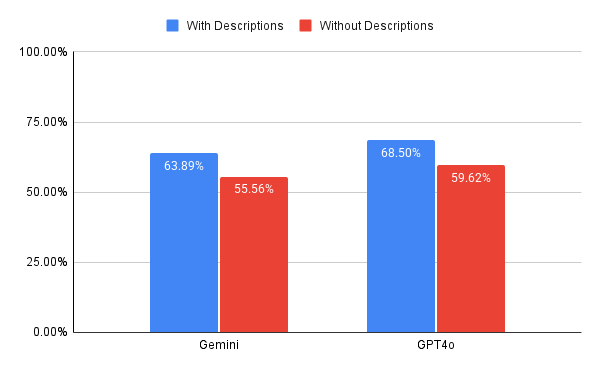}
    \caption{Correctness Accuracy (\%)}
    \label{fig:correctness_accuracy}
\end{figure}
With the mixed dataset, we ran the experiment with and without problem descriptions for both models. GPT4o outperformed Gemini in code correctness assessments with and without problem descriptions, as shown in Figure \ref{fig:correctness_accuracy}. When provided with descriptions, GPT4o was accurate 68.50\% of the time, compared to Gemini’s 63.89\%. 

\begin{figure}[h]
    \centering
    \includegraphics[width=1\linewidth]{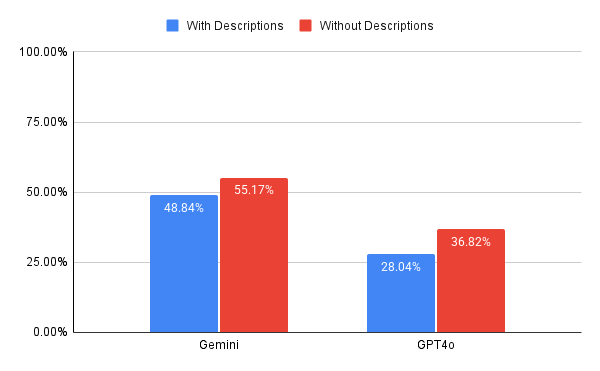}
    \caption{False Positive Rates (\%)}
    \label{fig:fp_rates}
\end{figure}
\begin{figure}[h]
    \centering
    \includegraphics[width=1\linewidth]{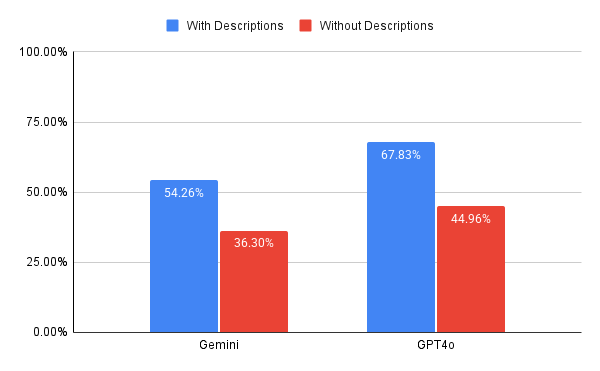}
    \caption{Correction Ratios (\%)}
    \label{fig:correction_ratios}
\end{figure}
\begin{figure}[h]
    \centering
    \includegraphics[width=1\linewidth]{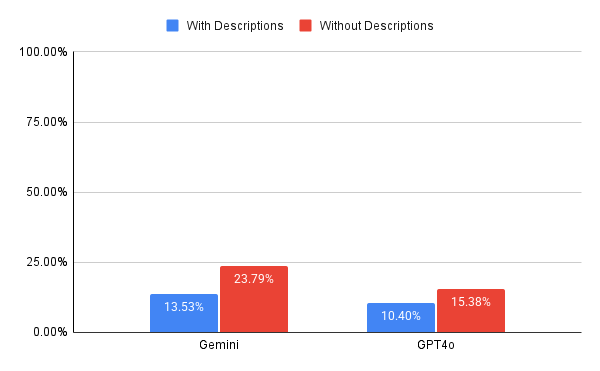}
    \caption{Regression Ratios (\%)}
    \label{fig:regression_ratios}
\end{figure}

Looking at the false positive rates in Figure \ref{fig:fp_rates}, we see that GPT4o is considerably better than Gemini. Both models perform poorer without the descriptions. 
\begin{figure}[h]
    \centering
    \includegraphics[width=1\linewidth]{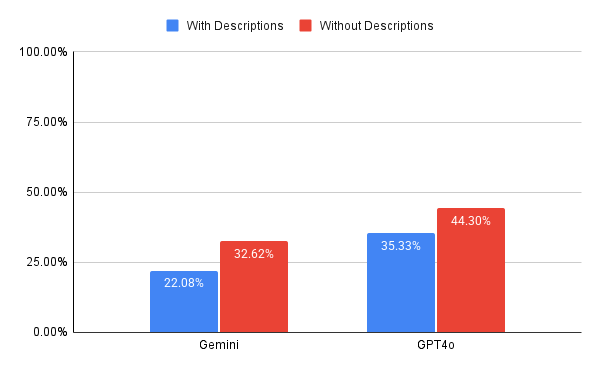}
    \caption{False Negative Rates (\%)}
    \label{fig:fn_rates}
\end{figure}
False negative rates in Figure \ref{fig:fn_rates} differ from the rest of the metrics because Gemini stands better than GPT4o. Although it may seem contradictory, it complements the false positive rates seen in Figure \ref{fig:fp_rates}. In terms of correction, GPT4o performed better than Gemini with and without problem descriptions, as shown in Figure \ref{fig:correction_ratios}. GPT4o had a higher correction ratio at 67.83\%, surpassing Gemini’s 54.26\%. Looking at the regression ratios, GPT4o was better in all configurations, as shown in Figure \ref{fig:regression_ratios}. GPT4o had a regression ratio of 10.43\%, while Gemini’s was 13.53\%.
Our experiments showed that both models performed poorer when the prompt did not provide the problem descriptions. This was especially apparent in regression and correction ratios, where differences of up to 22.87\% were observed.
\subsection{Ground Truth Dataset Experiment Results}
Using the ground truth dataset, we ran the experiment for both models with and without problem descriptions. We did not calculate correction ratios since all code blocks were already "Correct." 


\begin{figure}[h]
    \centering
    \includegraphics[width=1\linewidth]{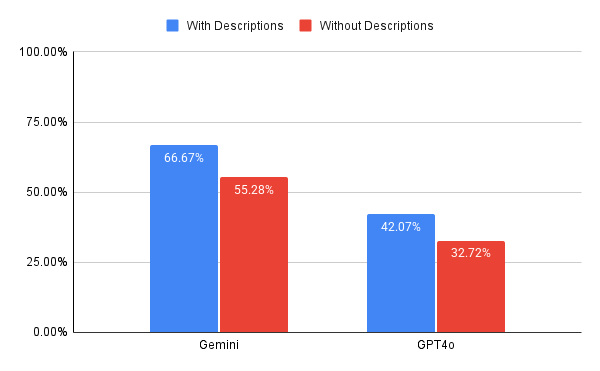}
    \caption{Correctness Accuracy (\%) with Ground Truth}
    \label{fig:gt_correctness_accuracy}
\end{figure}
The correctness accuracy results contradicted the mixed dataset findings as seen in Figure \ref{fig:gt_correctness_accuracy}. Gemini outperformed GPT4o with 66.67\%, compared to GPT4o's 42.07\% with problem descriptions. The regression ratio results were similar to the mixed dataset results with GPT4o outperforming Gemini with 9.96\% to Gemini's 12.40\% with problem descriptions. Both models performed poorer without problem descriptions, though with smaller differences of up to 12.40\%.
\begin{figure}[h]
    \centering
    \includegraphics[width=1\linewidth]{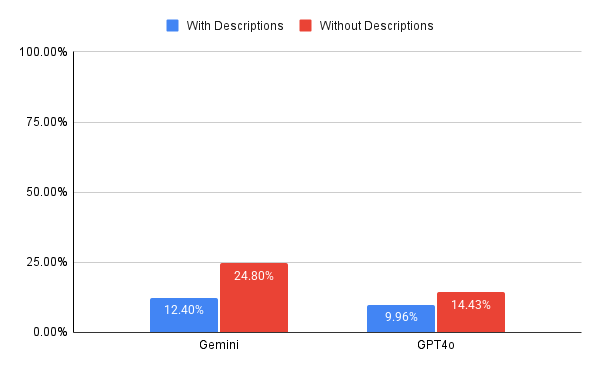}
    \caption{Regression Ratios (\%) with Ground Truth}
    \label{fig:gt_regression_ratios}
\end{figure}

\section{Discussion}
\label{discussion}

\subsection{Revisiting Research Question 1}
Our findings suggest that LLMs can evaluate code changes for approval or rejection with moderate accuracy. The highest value we observed was 68.50\% (GPT4o  with problem descriptions, mixed dataset).  In contrast, Gemini performed worse on the mixed dataset yet better on the ground truth dataset. This raises questions about whether the code type affects LLM performance. To mitigate such concerns, our experimental setup can help practitioners run tests using their own code, guiding them toward the best-performing LLM for their specific needs.

As shown in Figure \ref{fig:fp_rates} and Figure \ref{fig:fn_rates}, Gemini is more likely to misclassify "Incorrect" code blocks as "Correct," whereas GPT4o tends to make the opposite error more often. Given these scenarios, higher false negative rates are preferable to higher false positive rates because merging faulty code into the mainline can lead to quality issues, such as bugs. In contrast, the opposite error primarily inconveniences the author, which we consider to be a more minor issue by comparison. For this reason, we believe that the false positive and false negative rates are significant factors when choosing an LLM.

Finally, we consistently observed that LLMs perform better with problem descriptions. This suggests that practitioners adopting automated code review tools should ensure their code changes include clear, helpful comments. It should be noted that we do not refer to code comments specifically. Depending on the tool setup, other descriptions, such as pull request descriptions, can also serve the same purpose.

\subsection{Revisiting Research Question 2}
To effectively replace human code reviewers, LLMs need to provide effective code suggestions. In our experiments, we observed that LLMs can correct up to 67.83\% of incorrect code (GPT4o  with problem descriptions, mixed dataset). Our results suggest that code improvement suggestions of LLMs are moderately effective in improving code correctness. Without the problem descriptions, the results were consistently poorer.

In terms of regressions, we observed that up to 24.80\% of correct code blocks received incorrect code suggestions. The regression rates were higher, and even doubling, without the problem descriptions. The regression and correction ratio results also underlined the importance of comments. Overall, the key takeaway for us is that LLM code suggestions are not reliable enough for full automation. However, this does not mean they cannot be useful when used in moderation.  

\subsection{Human-in-the-loop LLM Code Review}

While our results were positive, they also show that LLMs exhibit significant error rates, with regression rates reaching up to 23.79\% and inaccurate approval decisions of 44.44\% (both from Gemini w/o problem descriptions, mixed dataset). Such errors could cause more harm than benefit, raising doubts about the reliability and accountability of fully automated LLM code reviews.  Furthermore, full automation overlooks crucial human-driven aspects of code review like knowledge transfer, team awareness, and shared code ownership \cite{Bacchelli2013_modernCodeReview}, \cite{Macleod2018_CRInTrenches}. Consequently, a hybrid process with human involvement is essential.
\begin{figure}[h]
    \centering
    \includegraphics[width=1\linewidth]{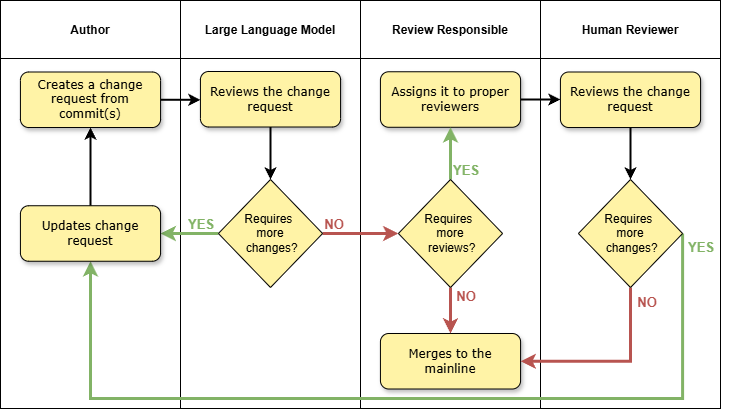}
    \caption{Human-in-the-loop LLM Code Review Process }
    \label{fig:combined_code_review}
\end{figure}

To address these shortcomings, we propose the "Human-in-the-loop LLM Code Review" process \ref{fig:combined_code_review}. To ensure accountability, the author, not the LLM, implements suggested changes. This LLM-author iteration continues until the LLM suggests no further modifications, with organizations setting iteration limits. Next, a "Review Responsible" determines if the LLM's review is sufficient or if additional human review is necessary, based on the change's criticality or complexity. If no further review is needed, the change is merged. Otherwise, human reviewers provide a second layer of oversight. If more changes are required, the process restarts; if not, the change is merged.  

This approach reduces manual effort while leveraging human expertise, facilitating knowledge transfer, and mitigating the regressions and faulty assessments observed in our experiments. The process is adaptable to organizational needs and aims to help practitioners establish their own LLM-assisted code review systems.

\section{Threats to Validity}
\label{threatstovalidity}
\subsection{Internal Validity}
Since we are experimenting with LLMs, the prompt plays a crucial role. We acknowledge that different prompts can yield different results, and ours followed a chain-of-thought approach. Our testing method for code suggestions is also subject to criticism, as it expects a code block rather than a textual suggestion. We chose to do it this way to ensure objective results. We extracted code from the YAML response and ran the corresponding unit tests. Since Python is indentation-sensitive, our prompts warned about indentation and correct YAML formatting. Overall, 94.70\% of the code was executed without errors, while 4.08\% had indentation errors and 1.08\% had YAML format errors. Because our prompt explicitly warned about these issues, we classified erroneous code blocks as "Incorrect," regardless of whether they would pass unit tests if the indentation or YAML errors were fixed.
\subsection{External Validity}
In this research, our scope is limited to Python. Therefore our findings are only directly generalizable to Python. Additionally, due to the stochastic nature of LLMs \cite{Minaee2024_LLMSurvey}, their outputs are not always identical. To account for this variability, we expanded our sample size by running our experiment three times and reporting the average results.
\subsection{Construct Validity}
Our evaluation was conducted on GPT-4o and Gemini, and other LLMs may exhibit different behaviors under the same conditions. The HumanEval dataset \cite{chen2021_HumanEvalDataset} consists of simple questions, while the mixed dataset was AI-generated, raising concerns about generalizability. We failed to find a dataset of human-generated code with unit tests with a similar diversity of correctness. To gain deeper and more reliable insights, future experiments should be conducted on code from real software projects. We provide practitioners with our experiment setup and source code, enabling them to generate their own results.

\subsection{Conclusion Validity}

In a practitioner setting, unit testing is not always conducted. This may affect the applicability of our results, as practitioners may apply different criteria for code approval or unit tests might not have enough coverage. We used unit tests since it was a reliable way to get objective results. Practitioners may create their own labels to run their own experiments. 

\section{Conclusion}
\label{conclusion}
In this study, we experimentally evaluated the code review capabilities of state-of-the-art LLMs, Gemini 2.0 Flash, and GPT4o. The LLMs were prompted to assess the correctness of code blocks and provide suggestions if needed. We observed that LLM performance consistently improved when a description of the code's intended functionality was provided. This finding highlights the importance of code comments and pull request descriptions for obtaining better reviews. The results across different datasets showed significant variations, highlighting the need for customized testing for different codebases. Practitioners can use our setup to identify the best-performing LLM and prompt tailored to their needs. 

The results indicated that the LLM's assessments had moderate accuracy, and its suggestions were moderately effective.  This suggests that a fully automated code review process may be unreliable. We suggested a hybrid process combining human involvement and LLMs to reduce manual effort and increase reliability while preserving the human-driven benefits of code review. We called it "Human-in-the-loop LLM Code Review." Together with our experimental setup, this process can help organizations establish their own LLM-based automated code review systems.

\section{Data and Code Availability}
\label{dataavailability}
The data obtained from this study, as well as the code used, are available in our Zenodo record at: https://doi.org/10.5281/zenodo.14962566

\bibliographystyle{plain}
\bibliography{refs}

\end{document}